\documentclass[usenatbib,a4paper]{mn2e}
\usepackage{graphicx}
\usepackage{pslatex}
\usepackage{color}
\usepackage{comment}
\title[Stellar assembly and AGN feedback at high redshift]{Constraining stellar assembly and AGN feedback at the peak epoch of star formation}
\author[T. Kimm et al.]{
\parbox[t]{\textwidth}{
T. Kimm$^{1}$\thanks{e-mail: taysun.kimm@astro.ox.ac.uk}, S.
Kaviraj$^{2,1}$, J. E. G. Devriendt$^{1,3}$, S. H. Cohen$^4$, R.
A. Windhorst$^4$, Y. Dubois$^{5,1}$, A. Slyz$^1$, N. P.
Hathi$^{6}$, R. E. Ryan Jr$^7$, R. W. O'Connell$^8$, M. A.
Dopita$^{9,10}$, and J. Silk$^5$ }
\vspace*{6pt} \\
$^1$ Department of Physics, Denys Wilkinson Building, Keble Road, Oxford, OX1 3RH, UK\\
$^2$ Blackett Laboratory, Imperial College London, London SW7 2A, UK\\
$^3$ CRAL, Universit\'e Claude Bernard Lyon I, CNRS UMR 5574, ENS-Lyon, 9 Avenue Charles Andr\'e, F-69561 Saint-Genis Laval, France\\
$^4$ School of Earth and Space Exploration, Arizona State University, Tempe, AZ 85287-1404, USA\\
$^5$ Institut d'Astrophysique de Paris, Universit\'e Pierre et Marie Curie Paris 6, CNRS, UMR 7095, F-75014, Paris, France \\
$^6$ Carnegie Observatories, 813 Santa Babara Street, Pasadena, California, 91101, USA\\
$^7$ Space Telescope Science Institute, 3700 San Martin Drive, Baltimore, MD 21218, USA\\
$^8$ Department of Astronomy, University of Virginia, P.O. Box 3818, Charlottesville, VA 22903, USA\\
$^9$ Research School of Physics and Astronomy, The Australian National University, ACT 2611, Australia\\
$^{10}$ Astronomy Department, King Abdulaziz University, P.O. Box 80203, Jeddah, Saudi Arabia
}
\begin{document}
\maketitle

\newcommand{\msun}{\mbox{$M_\odot$}}
\newcommand{\mvir}{\mbox{$M_{\rm vir}$}}
\newcommand{\rvir}{\mbox{$R_{\rm vir}$}}
\newcommand{\ramses}{\mbox{{\sc \small Ramses}}}
\newcommand{\nh}{\mbox{$n_{\rm H}$}}
\newcommand{\nth}{\mbox{$n_{\rm th}$}}
\newcommand{\mstar}{\mbox{$m_{\rm star}$}}

\newcommand{\blue}{\color{blue}}

\newcommand{\araa}{\mbox{ARA\&A}}
\newcommand{\aap}{\mbox{A\&A}}
\newcommand{\apj}{\mbox{ApJ}}
\newcommand{\aj}{\mbox{AJ}}
\newcommand{\apjl}{\mbox{ApJL}}
\newcommand{\apjs}{\mbox{ApJS}}
\newcommand{\mnras}{\mbox{MNRAS}}
\newcommand{\nat}{\mbox{Nature}}

\begin{abstract}
We study stellar assembly and feedback from active galactic nuclei
(AGN) around the epoch of peak star formation ($1\la z \la 2$), by
comparing hydrodynamic simulations to rest-frame $UV$-optical galaxy
colours from the Wide Field Camera 3 (WFC3) Early-Release Science
(ERS) Programme. Our Adaptive Mesh Refinement simulations include
metal-dependent radiative cooling, star formation, kinetic
outflows due to supernova explosions, and feedback from
supermassive black holes. Our model assumes that when gas accretes
onto black holes, a fraction of the energy is used to form either
thermal winds or sub-relativistic momentum-imparting collimated
jets, depending on the accretion rate. We find that the predicted
rest-frame $UV$-optical colours of galaxies in the model that
includes AGN feedback are in broad agreement with the observed
colours of the WFC3 ERS sample at $1\la z\la 2$. The predicted
number of massive galaxies also matches well with observations in
this redshift range. However, the massive galaxies are predicted
to show higher levels of residual star formation activity than the
observational estimates, suggesting the need for further
suppression of star formation without significantly altering the
stellar mass function. We discuss possible improvements, involving
faster stellar assembly through enhanced star formation during
galaxy mergers while star formation at the peak epoch is still
modulated by the AGN feedback.
\end{abstract}

\begin{keywords}
galaxies: formation -- galaxies: high-redshift  -- ultraviolet: galaxies
\end{keywords}

\voffset=-0.6in
\hoffset=0.1in

\section{Introduction}
The present-day galaxy luminosity function shows a rapid turnover
at luminosity $L_*$ \citep{cole01}, while $\Lambda$CDM cosmology
predicts a power-law form for the dark matter halo mass function
\citep{jenkins01}. The low mass-to-light ratio in low- and
high-mass halos is interpreted as evidence for the existence of
baryonic feedback, that regulates star formation
\citep[e.g.][]{binney04}. Consensus favours a picture in which
energy from active galactic nuclei (AGN) regulates the formation
of massive galaxies \citep[e.g.][]{ciotti97,kaviraj11b}, despite the
significant difference in the scale of a supermassive black hole
(SMBH) from that of its host galaxy. This picture appears to be
supported by the tight observed correlation between SMBH mass and
the bulge of its host galaxy \citep[e.g.][]{haring04},
evidence for the suppression of gas cooling and star formation due
to AGN interaction with the interstellar/intracluster medium
\citep[e.g.][]{mcnamara07}, high-velocity galactic
outflows that are commonly detected in quasars
\citep{pounds03,ganguly08} and X-ray cavities inflated by radio
jets, associated with AGN, that are observed in cluster centres
\citep{fabian06}. Nevertheless, it is still unclear as to how, to what
extent, and at what epochs, AGN feedback
 shapes the star formation history (SFH) of galaxies in the Universe.

Several studies have used semi-analytic or fully hydrodynamic
models of galaxy formation, to understand the effect of negative
feedback on the cosmic SFH \citep[e.g.][]{fontanot09}.
While detailed physical ingredients and prescriptions differ in
each model, such studies agree on the importance of AGN feedback
for reproducing the masses, colours, and star formation rates
(SFRs) of local massive galaxies \citep[e.g.][]{kimm09}. However,
the comparisons are almost exclusively performed in the nearby
Universe, while the bulk of the stellar assembly in these galaxies
takes place at high redshift ($z>1$). Unfortunately, 8-10 Gyrs of
evolution can easily wash out many of the {\it details} of stellar
assembly, and thus it is necessary to investigate the epoch {\it
at which the bulk of the star formation takes place ($1\la z \la
3$)} \citep{madau98}. Confronting the observed galaxy colours with
the models at these epochs represents a more stringent test of
their reliability.

In this \emph{Letter}, we study the stellar assembly in massive
galaxies, by comparing a fully hydrodynamical cosmological
simulation to \emph{rest-frame} UV-optical galaxy colours from the
WFC3 Early Release Science (ERS) programme \citep{windhorst11}.
The rest-frame UV (shortward of 3000\AA), which is sensitive to
even residual amounts of star formation, is a powerful tracer of
how quiescent a galaxy is \citep[e.g.][]{Kaviraj11,rutkowski12},
enabling us to constrain the quenching of star formation due to
AGN and stellar feedback. By contrast, the rest-frame optical
colours constrain the average epoch of stellar mass assembly. This
is the first direct comparison to simulations of rest-frame
UV-optical colours at the epoch of peak star formation. We
describe the data and simulations in \S~\ref{sec:obs} and
\ref{sec:sim} respectively. We present the comparison between
observations and models in \S~\ref{sec:res}, and discuss the
implications of our results on galaxy formation in
\S~\ref{sec:sum}. Throughout, we adopt
$(h_0,\,\Omega_m,\,\Omega_b,\,\Omega_{\Lambda},\,\sigma_8 ) =
(0.7,\,0.26,\,0.045,\,0.74,\,0.8)$ following the WMAP-3 results
\citep{hinshaw09}. All fluxes are based on the AB magnitude
system.

\vspace{-0.2in}

\section{Observations}
\label{sec:obs}
The WFC3 ERS programme \citep{windhorst11} has imaged $\sim$1/3 of
the GOODS-South field ($\sim$45 arcmin$^2$), using the WFC3 F225W,
F275W, F336W, F098M, F125W, and F160W filters, with exposure times
of 1-2 orbits per filter. In combination with the existing GOODS
BViz coverage \citep{Giavalisco2004}, this provides 10-filter
panchromatic coverage (0.2--1.7 $\mu m$) to point source depths of
AB $\sim$26.5--27.5 mag (UV--IR, 5$\sigma$). Here, we study
188 and 151 galaxies with $M_{\rm star}\ge 10^{10}\msun$, for
comparisons at $0.8\le z\le 1.2$ and $1.7\le z\le 2.3$
respectively. Note that the ERS sample is complete down to 
$\sim 10^9\msun$ and the observed images that trace both the rest frame UV and the 
rest frame optical are deep enough for all massive galaxies to be detected \citep{windhorst11}. 
For every galaxy, photometric 
redshifts were calculated using EAZY, 
based on all 10 filters with all possible combinations of pairs of 
the EAZY$_\_$v1.0 templates \citep{Brammer2008}.  
The absorption due to intervening intergalactic HI clouds is taken into consideration.
No redshift prior probability distribution specific to these fields is used. 
The resulting mean reduced $\chi^2$ of the fit is 0.94, and the typical 
redshift uncertainty of the sample is 
$\Delta z = \left|z_{\rm spec}-z_{\rm phot}\right| \simeq 0.15$. 
We note that no k-correction is applied to the observed sample, 
and that we refer the rest-frame UV and optical fluxes to the fluxes 
through the WFC3/ACS bandpass that {\em roughly} trace the 
GALEX $NUV$ or Johnson $B$/$V$. Thus our estimated rest-frame 
$NUV$ or optical fluxes are slightly different from GALEX $NUV$ or 
Johnson $B$/$V$ fluxes. 

In addition, the full suite of WFC3/ACS photometry of each galaxy is compared 
to a library of exponentially decaying model star formation histories, 
based on the \citet{bruzual03} stellar models. A wide range of ages (0.001--13 Gyr), 
decay timescales (0.1--9 Gyr), metallicities (0.005--2 $Z_\odot$) and 
dust extinctions $0 < E(B-V) < 2$ are employed, with the normalisation of the 
models yielding the stellar mass. The likelihood of individual models ($\chi^2/2$) are 
calculated and parameters are marginalised. The median of the  marginalised 
probability distributions are taken to be the best estimates, with the 25 and 75 
percentile values (which enclose 50\% of the total probability) providing an 
associated uncertainty \citep[see e.g.][]{Kaviraj11}. Specific SFRs (SSFRs) 
and stellar masses thus derived have uncertainties of $\sim$0.3 dex.


\section{Simulation}
\label{sec:sim}
In this paper, we use the octree-based Eulerian hydrodynamics
code, \ramses\ \citep{teyssier02}.
Metal-dependent radiative cooling is modelled based on
\citet{sutherland93}, and a uniform UV background is
instantaneously turned on at $z=10.5$ following \citet{haardt96}.
Star particles are created following a Schmidt law with 2\%
efficiency, when the density of a gas cell
reaches a critical density, $\rho_0=0.4\;{\rm H/cm}^3$. We use the
polytropic equation of state for the gas above the threshold
density, with the minimum temperature of $10^4\,{\rm K}$. Massive
stars are assumed to lose their mass through stellar winds and a
supernova phase, dispersing gas and metals (18\% and 2\% by mass,
respectively) into their surroundings. Feedback is modelled as an
isotropic kinetic outflow, which carries matter amounting to
ten times the mass loss from the stars  \citep{dubois08}.

Our model assumes that seed BHs of $10^5 \msun$ form in regions of
high gas density, ensuring that only one BH is formed per galaxy
\citep{dubois12}. The growth of the BH is tracked self-consistently, 
based on a modified Bondi-Holye-Lyttleton accretion rate. 
We adopt a density-dependent boost factor 
$\alpha_{\rm BH}=(\rho/\rho_0)^2$, following \citet{booth09}. 
When gas accretes onto BHs, we assume that a
central BH can impact the ambient gas in two ways, depending on
the Eddington ratio ($\chi\equiv \dot{m}/\dot{m}_{\rm edd}$). For
a high accretion rate ($\chi\ge0.01$), 1.5\% of the accretion
energy is injected as thermal energy (quasar-like mode), whereas
sub-relativistic ($v=10^4\;{\rm km/s}$) momentum-imparting
collimated winds are launched for a low accretion rate
($\chi<0.01$) event with 10\% efficiency \citep{dubois12}. The
parameters are chosen to match the stellar baryon fraction in
groups and clusters of galaxies (Dubois et al. {\sl in prep}). In
order to avoid artificial cooling and maximise the impact of AGN
feedback, we store the thermal energy and kinetic energy until it
either reaches $10^8$K or the energy corresponding to 10\% of the
BH mass. Therefore, the efficiency of the two modes should be
regarded as a rough estimate, not a true efficiency.

We have performed two cosmological simulations with the same
initial conditions generated by \texttt{mpgrafic}
\citep{prunet08}, a parallel version of the \texttt{grafic} package
\citep{bertschinger01}. The simulations contain 256$^3$ dark
matter particles with $m_{\rm dm}\simeq4\times10^9\msun$ in a 50
$h^{-1}$ Mpc comoving periodic box. We fix the maximum spatial
resolution to $\simeq0.38\;{\rm kpc/h}$ in physical units. The minimum
mass of the star particles is $1.65\times10^6\msun$.
The virial mass of the most massive halo in our simulated volume is 
$3.3\times10^{13}\msun$ at $z=1$. Note that the mass of dark matter halos 
used to compare with the observation in \S 4 is in the range 
$12 \la \log \mvir \la 13.5$, and hence our sample represents massive galaxies 
($\mstar\ge10^{11}\msun$ at $z=1$) forming in group/field environments. 
This allows for a fair comparison with the ERS sample, which is known to have 
few cluster galaxies \citep{windhorst11}. It is also worth mentioning that 36\% of 
the simulated massive galaxies are bulge-dominated at $z=1$, which seems 
compatible with the observed morphological fraction in the GOODS fields 
\citep{bundy05}.

\begin{figure*}
\centering
   \includegraphics[width=7.5cm]{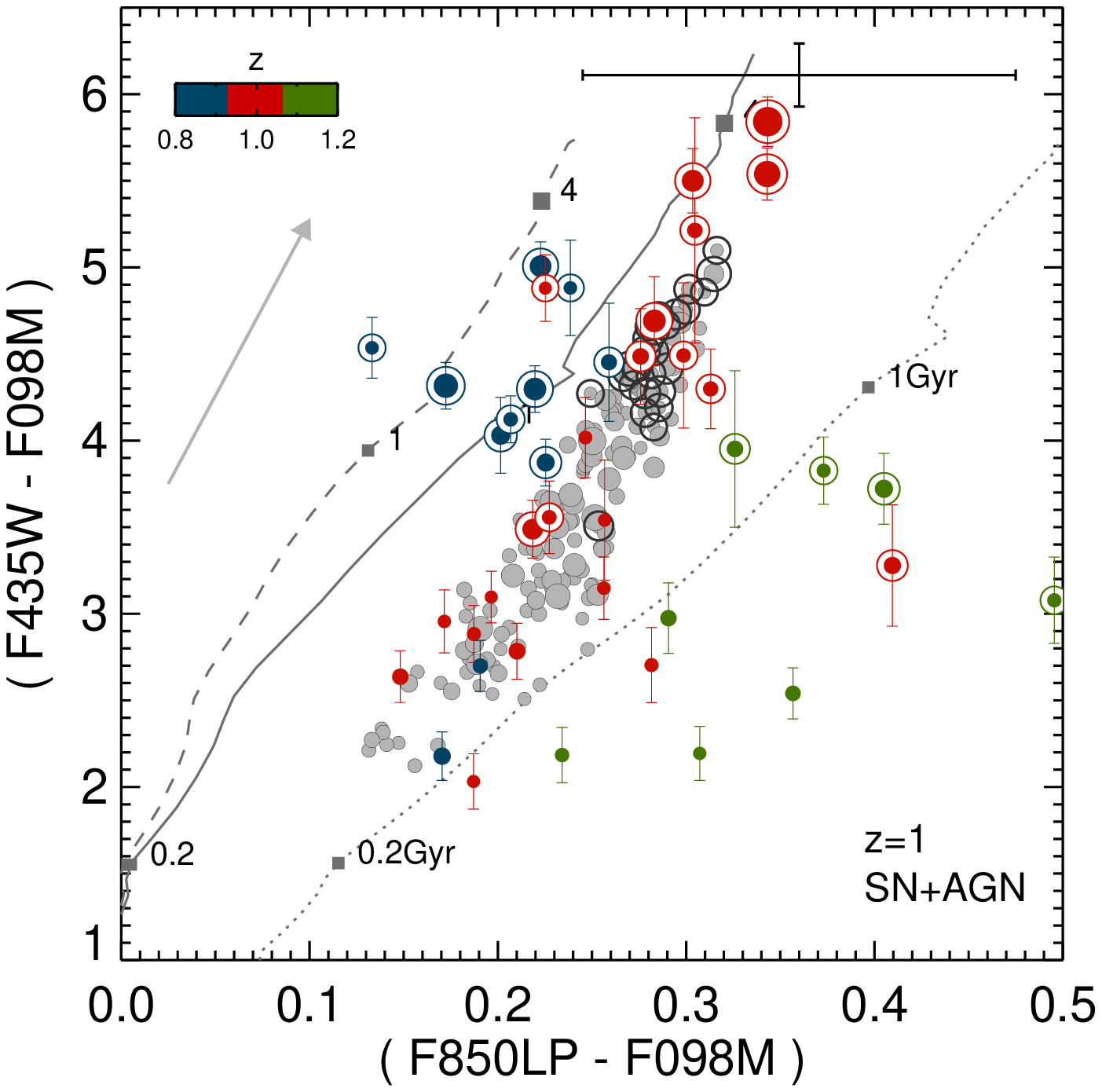}
   \includegraphics[width=7.5cm]{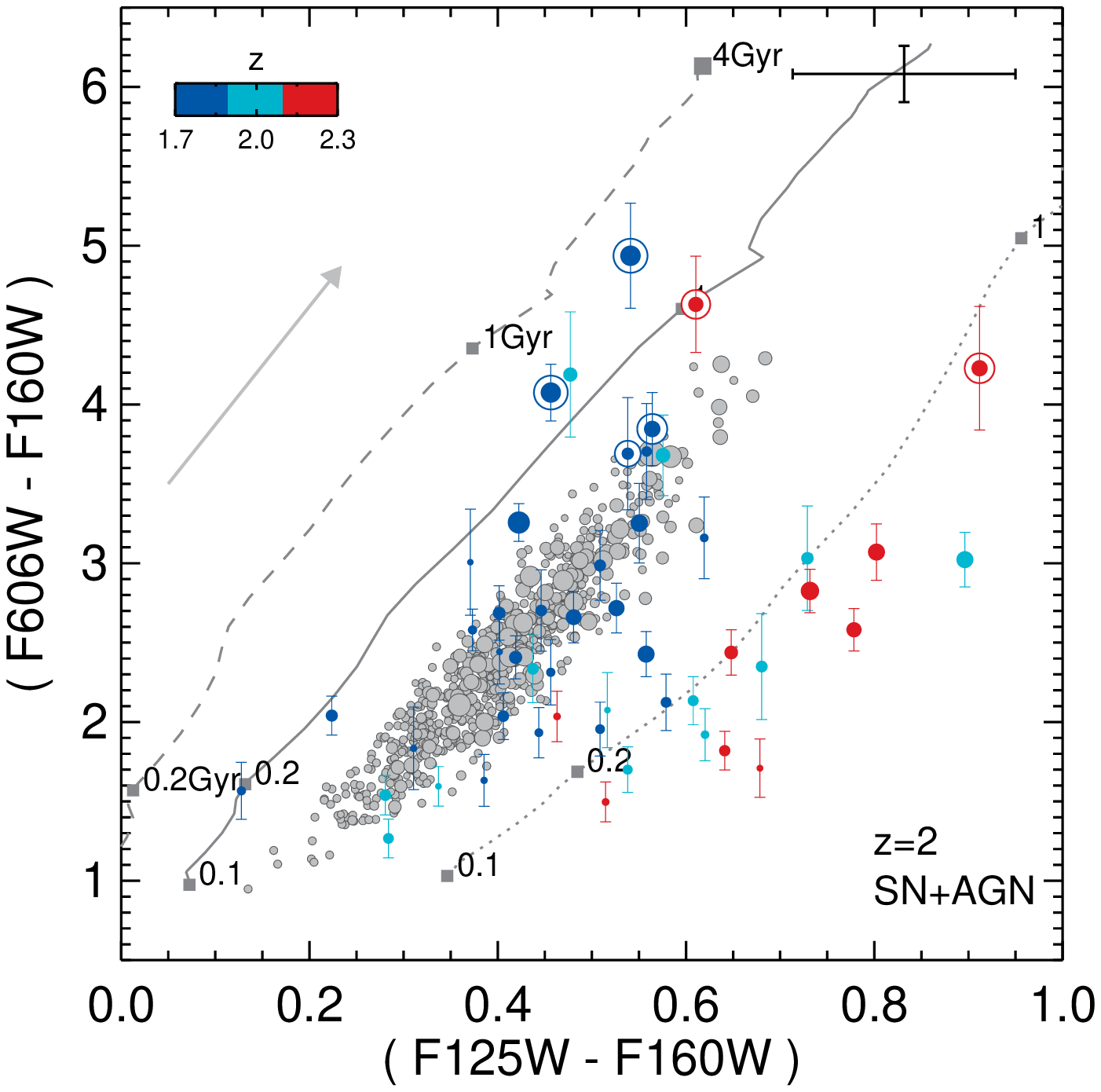}
   \caption{Colour comparisons of massive galaxies at $z=1$ (left; $m_{\rm
   star}\ge10^{11}\,\msun$) and $z=2$ (right; $m_{\rm star}\ga10^{10.5}\,\msun$) with 
   simulations that include AGN+supernova feedback. The observed colours on 
   the x and y axes trace rest-frame ($B-V$) and ($NUV-V$), respectively 
   (see text for more details). Filled circles with error bars indicate 
   the WFC3 ERS data,    while simulated galaxies are shown using filled grey circles. 
   The size of the symbol scales with stellar mass. The three lines represent 
   the expected colours of simple stellar populations (SSPs) with different ages at
   different redshift: in the left-hand panel, $z=$0.8 (dashed), 1.0 (solid), and 1.2 (dotted), 
   in the right-hand panel: $z=$1.7 (dashed), 2.0 (solid), and 2.3 (dotted). 
   For clarity, the colours of SSPs with different ages (0.1, 0.2, 1, and 4 Gyr) are 
   indicated by filled square symbols.   Points with a concentric circle denote galaxies 
   with low specific star formation rate (SSFR$<0.01\; {\rm Gyr}^{-1}$).   We indicate the 
   median uncertainty of the observed colours in the top right and the typical extinction 
   vectors of  the simulated galaxies in the left of each panel.}
   \label{fig:colour}
\end{figure*}

Predicted colours of simulated galaxies are generated by folding
the stellar spectra of individual star particles with the
stellar population models of \citet{bruzual03}. The optical depth
is calculated following the empirical calibration of
\citet{guiderdoni87}, and dust attenuation is calculated using the
\cite{cardelli89} extinction curve \citep[][Eq.~1]{devriendt10}.
We emphasise that {\it the attenuated spectra of the simulated galaxies
are redshifted to either $z=1$ or $z=2$ before convolving them
with the throughput of WFC3/ACS filters} to directly compare with
ERS galaxies.

\vspace{-0.2in}

\section{Results}
\label{sec:res}

In Fig.~\ref{fig:colour} we present the rest-frame UV and optical
colours of ERS galaxies around $z\simeq1$. We use the ACS F435W,
F850LP, and WFC3 F098M filters to trace the {\em rest-frame} $NUV$
($\sim$2300\AA), $B$, and $V$ bands, respectively. In order to test the
influence of AGN feedback, we restrict our sample to galaxies more
massive than $m_{\rm star}\ge 10^{11}\,\msun$ because SN feedback is 
ineffective in such large potential wells, leaving AGN feedback as the most 
plausible mechanism for modulating star formation. Three reference lines 
are included in the figure to help distinguish the age and redshift sequence. 
Each line corresponds to a simple stellar population (SSP) with various ages 
(marked next to the line). Differences between the lines indicate the change 
in the colours when a spectrum is redshifted to different values before it is
convolved with the filter throughputs ($z=$ 0.8, 1.0, and 1.2 [left panel] or 
$z=$ 1.7, 2.0, and 2.3 [right panel]). Observed galaxies show a large scatter 
in $(NUV-V)$ and $(B-V)$ colours, partly because no k-correction is applied.
When the redshift interval of the sample 
is constrained around $z\simeq1$, a colour sequence is apparent (red points), with
massive galaxies being redder in both UV and optical colours.

To make a detailed comparison, we study the SSFRs derived from SED
fitting (see Section 2), marking `quiescent' (SSFR
$\equiv\dot{m}_{\rm star}/m_{\rm star} \le 0.01 \, {\rm
Gyr^{-1}}$) galaxies in our sample with a concentric circle
in Fig.~\ref{fig:colour}. We find that almost all massive galaxies
with $(NUV-V) \ga 4$ mag are quiescent. It is worth emphasising
here that this provides a stringent test for any cosmological
galaxy formation model, since such low SSFRs are comparable to
those of nearby `red and dead' galaxies \citep[e.g.][]{kimm09}.
\emph{Thus, any process that shapes the star formation history of
massive galaxies must be powerful enough to quench star formation
at this early epoch.}

\begin{figure}
\centering
  \includegraphics[width=7.5cm]{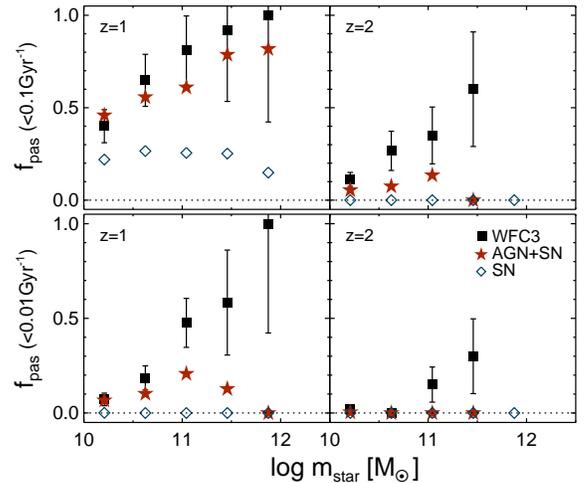}
  \caption{The fraction of passive galaxies for the observations (black squares), the model with supernova and AGN (red stars), and the model with supernova (blue diamonds). Top panels show the fraction of galaxies
  showing SSFR$<0.1\;{\rm Gyr^{-1}}$, while bottom panels display the fraction of galaxies which are an order
  of magnitude less active SSFR$<0.01\;{\rm Gyr^{-1}}$. Poisson errors are indicated by error bars.
  }
 \label{fig:ssfr}
\end{figure}

We find that the simulated galaxies with AGN feedback agree 
reasonably well with the observed sequence at
$z=1$. The left-hand panel of Fig.~\ref{fig:colour} indicates that
a majority of model galaxies with $(NUV-V) \ga4$ mag are indeed
quiescent (double circles), but not due to dust reddening, indicating that AGN
feedback is a plausible explanation for the emergence of quiescent
galaxies at $z\simeq1$. However, in the absence of AGN feedback
almost all model galaxies are predicted to be actively star
forming (Fig.~\ref{fig:ssfr}), suggesting that gravitational
heating by substructure alone \citep[e.g.][]{khochfar08} is not able to
account for the build-up of quiescent galaxies in group/field environments at this epoch
\citep[see also][]{johansson12}. Note that such interactions between infall of
substructure and the interstellar/intracluster medium is automatically accounted for
in our hydrodynamic calculations. We also see that the
model with AGN feedback reproduces the observed trend of blue
galaxies ($(NUV-V)\la3$ mag) being less massive. However, there is
an indication that massive galaxies with intermediate colours
($(NUV-V)\simeq$ 3.5 mag) are more frequent than observed in the
data (see the left panel in Fig.~\ref{fig:colour}).

In Fig.~\ref{fig:colour} (right panel), we show the corresponding
colour comparisons at $z\simeq2$. We only show galaxies with
masses greater than $3\times 10^{10}\msun$, which form more than
90\% of the progenitors of our model galaxy sample at $z=1$. Since
the redshift of the galaxy sample has changed, we now use the ACS
F606W and the WFC3 F125W and F160W filters to trace the rest-frame
$NUV$, $B$, and $V$ bands respectively. 
We find that our simulated galaxies at $z\simeq2$
compare better with WFC3 galaxies at $1.7\le z<1.9$ rather than
$1.9\le z<2.1$. We suspect that this is due, in part, to a
systematic underestimation of the photometric redshifts at these
epochs \citep[see also][]{williams09}. Note that the observed
galaxies at $1.7\le z<1.9$ are comparable to the model galaxies at
$z=2$, given the typical uncertainty of EAZY redshifts at these
epochs ($\Delta z \sim 0.15$). At $z\simeq2$ we find that there
exists a non-negligible number of massive galaxies with a very low
level of star formation activity, which are absent in the run with
AGN feedback. The massive galaxies ($\mstar\ga10^{11}\msun$) in
the model are actively star forming (SSFR$\sim{\rm 0.3\;
Gyr}^{-1}$), leading to colours ($(NUV-V)\la3$) that are
inconsistent with the observations.

\begin{figure}
\centering
   \includegraphics[width=7.5cm]{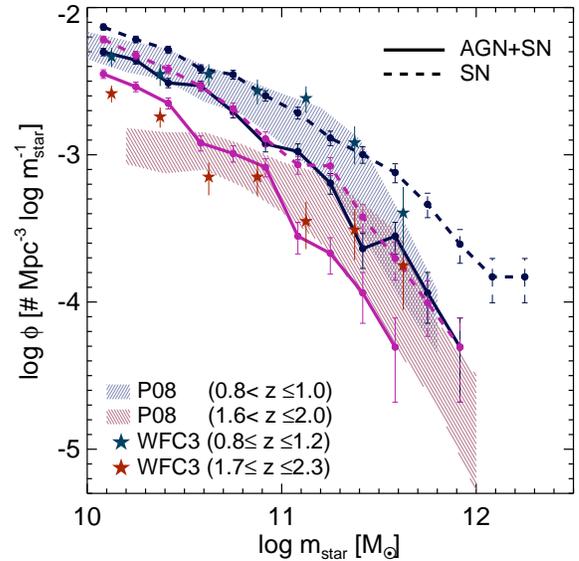}
   \caption{
   The stellar mass functions of galaxies at z$=$2 (reddish colours) and z$=$1 (bluish colours).
   Our estimations at each redshift intervals for simulated galaxies with and without AGN feedback are
   plotted as solid and dashed lines, respectively. Error bars indicate Poisson errors in the counts.
   For comparison, we display the mass functions at $0.8<z\le1.0$ (blue shading) and $1.6<z\le2.0$ (red shading)
   from \citet{perez-gonzalez08}. Also included as stars are the simple number counts from the WFC3 sample.
   }
   \label{fig:MF}
\end{figure}

Fig.~\ref{fig:ssfr} shows the fraction of passive galaxies,
which confirms that the modelled recent star formation needs to
be further suppressed in massive galaxies at  $z\simeq2$. While
our simulation predicts that $\la$ 20\,\% of massive
galaxies have SSFR$<0.1\;{\rm Gyr^{-1}}$ at $z\simeq2$, the
observations suggest that 30--60\% of the massive galaxy
population is passive (top right). At $z\simeq1$, a majority
(60--80\%) of the massive galaxies in the model have become
passive (top left), but when a more conservative definition for a
passive galaxy is used (SSFR$<0.01\;{\rm Gyr^{-1}}$), we find that
the passive fraction in the model drops to approximately 0--20\%
(bottom left). This is inconsistent with the observations, which
suggest that 50--100\% of massive galaxies are quiescent at this
epoch.

Nevertheless, the impact of AGN feedback becomes clearly visible
in the stellar mass functions (see Fig.~\ref{fig:MF}). The number of
massive galaxies in the AGN run ($m_{\rm star}\ga 10^{11}\msun$;
solid lines) are systematically smaller than the run without AGN
(dashed lines) by a factor of $\sim$ 3. Comparing twin halos that share position and mass 
in the feedback/no feedback runs, we find that the stellar masses of the galaxies they host are reduced by a
factor of 2--5, meaning that the mass functions are shifted
horizontally (to the left) due to AGN feedback. The mass functions
are in reasonable agreement with the stellar mass function estimated by \citet{perez-gonzalez08}
(red and blue shadings), and also compatible with the simple number counts
from the WFC3 galaxies at $1.7\le z \le 2.3$ (orange stars) or $0.8\le z \le 1.2$ (blue stars).

One may wonder whether our AGN feedback recipe is too strong,
given the apparent deficit of the most massive galaxies. However,
this can be partly attributed to the absence of density
fluctuations on scales comparable to the size of the simulation
\citep{sirko05,gnedin11}. Moreover, the fact that we need further
quenching of residual star formation in massive galaxies does not
appear to favour moderating the impact of AGN feedback. We discuss
this issue further in the next section.

\begin{figure}
\centering
   \includegraphics[width=7.2cm]{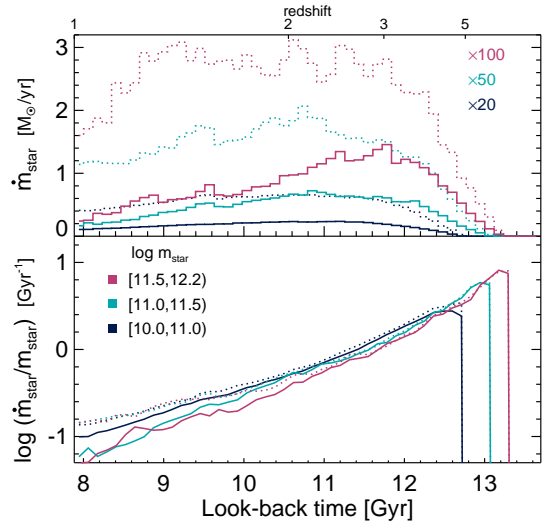}
   \caption{The median star formation history of simulated galaxies at $z=1$ in the run with (solid) and without (dotted) AGN feedback.
   {\it Top}: the median star formation rates of galaxies in three different mass bins.
   For clarity, the SFRs are scaled down by 100 for galaxies with $10^{11.5}\le m_{\rm star}<10^{12.2}$ (red),
   50 for galaxies with $10^{11.0}\le m_{\rm star}<10^{11.5}$ (cyan), and 20 for galaxies with $10^{10}\le m_{\rm star}<10^{11}$ (blue).
   {\it Bottom}: The specific star formation rates (SSFRs) as a function of look-back time.}
   \label{fig:sfh}
\end{figure}

Finally, in agreement with the downsizing picture \citep{cowie96},
we find that in the run with AGN feedback, the star formation
histories of more massive galaxies are peaked at higher redshift
(solid lines in the upper panel of Fig.~\ref{fig:sfh}). This can
also be seen in the bottom panel, where we plot the median SSFR of
1127, 117, and 29 galaxies in three mass bins, $10^{10}\le m_{\rm
star}<10^{11}\,\msun$, $10^{11}\le m_{\rm star}<10^{11.5}\,\msun$, and
$10^{11.5}\le m_{\rm star}<10^{12.2}\,\msun$, as a function of
look-back time. If we compare the galaxy SFHs in the runs with
(solid lines in the top panel) and without AGN feedback (dotted lines), we see that star
formation is preferentially suppressed at $1\la z \la 3$ at which
the bulk of the star formation takes place \citep{madau98}.

Interestingly, we see that the most massive galaxies have an
extended SFH, which may, at first sight, seem inconsistent with
the high observed [$\alpha$/Fe] ratios in local systems
\citep[e.g.][]{thomas05}. However, we note that it is the star
formation timescale in the \emph{subunits} that finally merge to
form the larger galaxy that determines its average [$\alpha$/Fe].
In other words it is not possible to directly infer the
[$\alpha$/Fe] from a `composite' SFH of the various subunits (such
as the one shown in Fig.~\ref{fig:sfh}), because much of the
broadness of the SFH is driven by the fact that star formation
peaks at different epochs in different subunits. Hence, the degree
of the discrepancy may not be substantial as it appears, and will
be tested with more detailed chemo-hydrodynamic simulations in a
forthcoming paper.

\vspace{-0.2in}

\section{Discussion and Conclusions}
\label{sec:sum}

 We have presented the first direct comparison of the rest-frame UV-optical
 colours of massive galaxies at $1\la z \la 2$ to the predictions of hydrodynamic
 simulations that include AGN feedback. We have shown that massive galaxies in
 the simulations are actively star forming at this redshift in the absence of a strong source
 of energetic feedback.
 In an attempt
 to estimate the stellar assembly of massive galaxies, we have incorporated a simple, jet/quasar-mode
 AGN feedback \citep{dubois12}, and found that it reproduces a reasonable fraction of
 {\em moderately} quiescent galaxies in both rest-frame UV and optical colours at $z=1$.
 However, the model slightly overpredicts the level of recent star formation activity in massive
 galaxies at $z=1$ and 2, compared with the observations. Despite this, the stellar mass functions
 in the model with AGN feedback agree reasonably well with the WFC3 ERS data
 (within the statistical uncertainties), due to the efficient quenching of star formation
 during the epoch of peak star formation.

The lack of massive, quiescent galaxies in the model seems to
require \emph{further} suppression of residual star formation
without significantly altering the stellar mass function. This
suggests that the peak epoch of the stellar assembly in massive
galaxies should commence earlier than we have predicted, while the
residual star formation must decline more sharply over time.
Although further star formation quenching is attainable by
introducing more enhanced feedback or other processes that we have
ignored (e.g. photoionisation of gas coolants due to soft X-ray
and extreme UV photons from stars, \citealt{cantalupo10}), changing
the peak epoch of stellar assembly, while keeping the stellar mass
function fixed may require well-balanced interplay between
feedback and star formation.

Regarding the extended nature of the predicted SFHs, a different
approach, involving a higher level of star formation activity in
the very early universe \citep[e.g.][]{stark09} is also appealing.
The advantages of this idea are 1) it does not violate the stellar
mass functions at $z\leq2$, 2) it alleviates the discrepancy in
[$\alpha$/Fe] ratio in massive galaxies to some degree
\citep[e.g.][]{pipino09}, although other routes, such as the use
of a top-heavy IMF may offer more promising solutions
\citep{arrigoni10}), and 3) star formation can be efficiently
suppressed with the same amount of energy released from the black
holes by consuming gas earlier on. Even though the physical origin
of such high specific star formation rates is unclear, we note
that proper modelling of gas fragmentation in galaxy mergers can
easily elevate star formation by an order of magnitude
\citep{teyssier10}, and may even be able to match the observed high
SSFRs in the early universe \citep[e.g.][]{khochfar11}.

We also note that the mass and spatial resolution of a simulation plays a role in determining
the star formation history of the model galaxies. Adopting a
better resolution allows us to resolve smaller halos in the early
universe ($z\ga5$), leading to earlier star formation, with an
order of magnitude enhancement in the star formation activity at
this epoch \citep{rasera06}. This will naturally increase the mean
age of the massive galaxies \citep[e.g.][]{johansson12}. However,
the question is how many stars form during this epoch, compared 
to the total stellar mass at $z\simeq1$ or 2. 
Inspection of the top panel of Fig.~\ref{fig:sfh} shows that an order of
magnitude increase in the SFR at $z\ge 5$ will not significantly
affect the overall SFHs.

Given the simplicity of our modelling of AGN feedback, the
agreement between observations and models is encouraging. An
important challenge will be to match chemical abundance ratios in
massive galaxies, while producing the correct number of red and
dead galaxies at high redshift. In a forthcoming paper we will
investigate these issues in detail, focussing on more realistic
modelling of the starburst phases and the interplay between
feedback and star formation at high redshift.

\vspace{-0.2in}

\section*{Acknowledgements}
The simulations were carried out on the DiRAC facility, funded by
STFC, the Large Facilities Capital Fund of BIS and the University
of Oxford. TK acknowledges support from a Clarendon DPhil
studentship. JD and AS's research is supported by Adrian Beecroft,
the Oxford Martin School and the STFC.
This paper is based on Early Release Science observations made by the
WFC3 Scientific Oversight Committee. We are grateful to the Director of the
Space Telescope Science Institute, Dr. Matt Mountain, for generously awarding
Director's Discretionary time for this program.
\vspace{-0.2in}

\small
\bibliographystyle{mn2e}
\bibliography{refs}

\begin{thebibliography}{47}
\expandafter\ifx\csname natexlab\endcsname\relax\def\natexlab#1{#1}\fi

\bibitem[{{Arrigoni} {et~al}\mbox{.}(2010){Arrigoni}, {Trager}, {Somerville},
  \& {Gibson}}]{arrigoni10}
{Arrigoni} M., {Trager} S.~C., {Somerville} R.~S., {Gibson} B.~K., 2010,
  \mnras, 402, 173

\bibitem[{{Bertschinger}(2001)}]{bertschinger01}
{Bertschinger} E., 2001, \apjs, 137, 1

\bibitem[{{Binney}(2004)}]{binney04}
{Binney} J., 2004, \mnras, 347, 1093

\bibitem[{{Booth} \& {Schaye}(2009)}]{booth09}
{Booth} C.~M., {Schaye} J., 2009, \mnras, 398, 53

\bibitem[{{Brammer} {et~al}\mbox{.}(2008){Brammer}, {van Dokkum}, \&
  {Coppi}}]{Brammer2008}
{Brammer} G.~B., {van Dokkum} P.~G., {Coppi} P., 2008, \apj, 686, 1503

\bibitem[{{Bruzual} \& {Charlot}(2003)}]{bruzual03}
{Bruzual} G., {Charlot} S., 2003, \mnras, 344, 1000

\bibitem[{{Bundy} {et~al}\mbox{.}(2005){Bundy}, {Ellis}, \&
  {Conselice}}]{bundy05}
{Bundy} K., {Ellis} R.~S., {Conselice} C.~J., 2005, \apj, 625, 621

\bibitem[{{Cantalupo}(2010)}]{cantalupo10}
{Cantalupo} S., 2010, \mnras, 403, L16

\bibitem[{{Cardelli} {et~al}\mbox{.}(1989){Cardelli}, {Clayton}, \&
  {Mathis}}]{cardelli89}
{Cardelli} J.~A., {Clayton} G.~C., {Mathis} J.~S., 1989, \apj, 345, 245

\bibitem[{{Ciotti} \& {Ostriker}(1997)}]{ciotti97}
{Ciotti} L., {Ostriker} J.~P., 1997, \apjl, 487, L105+

\bibitem[{{Cole} {et~al}\mbox{.}(2001){Cole}, {Norberg}, {Baugh}, {Frenk},
  {Bland-Hawthorn}, {Bridges}, {Cannon}, {Colless}, {Collins}, {Couch},
  {Cross}, {Dalton}, {De Propris}, {Driver}, {Efstathiou}, {Ellis},
  {Glazebrook}, {Jackson}, {Lahav}, {Lewis}, {Lumsden}, {Maddox}, {Madgwick},
  {Peacock}, {Peterson}, {Sutherland}, \& {Taylor}}]{cole01}
{Cole} S. {et~al.}, 2001, \mnras, 326, 255

\bibitem[{{Cowie} {et~al}\mbox{.}(1996){Cowie}, {Songaila}, {Hu}, \&
  {Cohen}}]{cowie96}
{Cowie} L.~L., {Songaila} A., {Hu} E.~M., {Cohen} J.~G., 1996, \aj, 112, 839

\bibitem[{{Devriendt} {et~al}\mbox{.}(2010){Devriendt}, {Rimes}, {Pichon},
  {Teyssier}, {Le Borgne}, {Aubert}, {Audit}, {Colombi}, {Courty}, {Dubois},
  {Prunet}, {Rasera}, {Slyz}, \& {Tweed}}]{devriendt10}
{Devriendt} J. {et~al.}, 2010, \mnras, 403, L84

\bibitem[{{Dubois} {et~al}\mbox{.}(2012){Dubois}, {Devriendt}, {Slyz}, \&
  {Teyssier}}]{dubois12}
{Dubois} Y., {Devriendt} J., {Slyz} A., {Teyssier} R., 2012, \mnras, 420, 2662

\bibitem[{Dubois \& Teyssier(2008)}]{dubois08}
Dubois Y., Teyssier R., 2008, \aap, 477, 79

\bibitem[{{Fabian} {et~al}\mbox{.}(2006){Fabian}, {Sanders}, {Taylor}, {Allen},
  {Crawford}, {Johnstone}, \& {Iwasawa}}]{fabian06}
{Fabian} A.~C., {Sanders} J.~S., {Taylor} G.~B., {Allen} S.~W., {Crawford}
  C.~S., {Johnstone} R.~M., {Iwasawa} K., 2006, \mnras, 366, 417

\bibitem[{{Fontanot} {et~al}\mbox{.}(2009){Fontanot}, {De Lucia}, {Monaco},
  {Somerville}, \& {Santini}}]{fontanot09}
{Fontanot} F., {De Lucia} G., {Monaco} P., {Somerville} R.~S., {Santini} P.,
  2009, \mnras, 397, 1776

\bibitem[{{Ganguly} \& {Brotherton}(2008)}]{ganguly08}
{Ganguly} R., {Brotherton} M.~S., 2008, \apj, 672, 102

\bibitem[{{Giavalisco} {et~al}\mbox{.}(2004){Giavalisco}, {}, \& {et
  al.}}]{Giavalisco2004}
{Giavalisco} M., {}, {et al.}, 2004, \apjl, 600, L93

\bibitem[{{Gnedin} {et~al}\mbox{.}(2011){Gnedin}, {Kravtsov}, \&
  {Rudd}}]{gnedin11}
{Gnedin} N.~Y., {Kravtsov} A.~V., {Rudd} D.~H., 2011, \apjs, 194, 46

\bibitem[{{Guiderdoni} \& {Rocca-Volmerange}(1987)}]{guiderdoni87}
{Guiderdoni} B., {Rocca-Volmerange} B., 1987, \aap, 186, 1

\bibitem[{{Haardt} \& {Madau}(1996)}]{haardt96}
{Haardt} F., {Madau} P., 1996, \apj, 461, 20

\bibitem[{{H{\"a}ring} \& {Rix}(2004)}]{haring04}
{H{\"a}ring} N., {Rix} H.-W., 2004, \apjl, 604, L89

\bibitem[{{Hinshaw} {et~al}\mbox{.}(2009){Hinshaw}, {Weiland}, {Hill},
  {Odegard}, {Larson}, {Bennett}, {Dunkley}, {Gold}, {Greason}, {Jarosik},
  {Komatsu}, {Nolta}, {Page}, {Spergel}, {Wollack}, {Halpern}, {Kogut},
  {Limon}, {Meyer}, {Tucker}, \& {Wright}}]{hinshaw09}
{Hinshaw} G. {et~al.}, 2009, \apjs, 180, 225

\bibitem[{{Jenkins} {et~al}\mbox{.}(2001){Jenkins}, {Frenk}, {White},
  {Colberg}, {Cole}, {Evrard}, {Couchman}, \& {Yoshida}}]{jenkins01}
{Jenkins} A., {Frenk} C.~S., {White} S.~D.~M., {Colberg} J.~M., {Cole} S.,
  {Evrard} A.~E., {Couchman} H.~M.~P., {Yoshida} N., 2001, \mnras, 321, 372

\bibitem[{{Johansson} {et~al}\mbox{.}(2012){Johansson}, {Naab}, \&
  {Ostriker}}]{johansson12}
{Johansson} P.~H., {Naab} T., {Ostriker} J.~P., 2012, astro-ph/1202.3441J

\bibitem[{{Kaviraj} {et~al}\mbox{.}(2011{\natexlab{a}}){Kaviraj}, {Schawinski},
  {Silk}, \& {Shabala}}]{kaviraj11b}
{Kaviraj} S., {Schawinski} K., {Silk} J., {Shabala} S.~S., 2011{\natexlab{a}},
  \mnras, 415, 3798

\bibitem[{{Kaviraj} {et~al}\mbox{.}(2011{\natexlab{b}}){Kaviraj}, {Tan},
  {Ellis}, \& {Silk}}]{Kaviraj11}
{Kaviraj} S., {Tan} K.-M., {Ellis} R.~S., {Silk} J., 2011{\natexlab{b}},
  \mnras, 411, 2148

\bibitem[{{Khochfar} \& {Ostriker}(2008)}]{khochfar08}
{Khochfar} S., {Ostriker} J.~P., 2008, \apj, 680, 54

\bibitem[{{Khochfar} \& {Silk}(2011)}]{khochfar11}
{Khochfar} S., {Silk} J., 2011, \mnras, 410, L42

\bibitem[{{Kimm} {et~al}\mbox{.}(2009){Kimm}, {Somerville}, {Yi}, {van den
  Bosch}, {Salim}, {Fontanot}, {Monaco}, {Mo}, {Pasquali}, {Rich}, \&
  {Yang}}]{kimm09}
{Kimm} T. {et~al.}, 2009, \mnras, 394, 1131

\bibitem[{{Madau} {et~al}\mbox{.}(1998){Madau}, {Pozzetti}, \&
  {Dickinson}}]{madau98}
{Madau} P., {Pozzetti} L., {Dickinson} M., 1998, \apj, 498, 106

\bibitem[{{McNamara} \& {Nulsen}(2007)}]{mcnamara07}
{McNamara} B.~R., {Nulsen} P.~E.~J., 2007, \araa, 45, 117

\bibitem[{{P{\'e}rez-Gonz{\'a}lez}
  {et~al}\mbox{.}(2008){P{\'e}rez-Gonz{\'a}lez}, {Rieke}, {Villar}, {Barro},
  {Blaylock}, {Egami}, {Gallego}, {Gil de Paz}, {Pascual}, {Zamorano}, \&
  {Donley}}]{perez-gonzalez08}
{P{\'e}rez-Gonz{\'a}lez} P.~G. {et~al.}, 2008, \apj, 675, 234

\bibitem[{{Pipino} {et~al}\mbox{.}(2009){Pipino}, {Devriendt}, {Thomas},
  {Silk}, \& {Kaviraj}}]{pipino09}
{Pipino} A., {Devriendt} J.~E.~G., {Thomas} D., {Silk} J., {Kaviraj} S., 2009,
  \aap, 505, 1075

\bibitem[{{Pounds} {et~al}\mbox{.}(2003){Pounds}, {Reeves}, {King}, {Page},
  {O'Brien}, \& {Turner}}]{pounds03}
{Pounds} K.~A., {Reeves} J.~N., {King} A.~R., {Page} K.~L., {O'Brien} P.~T.,
  {Turner} M.~J.~L., 2003, \mnras, 345, 705

\bibitem[{{Prunet} {et~al}\mbox{.}(2008){Prunet}, {Pichon}, {Aubert},
  {Pogosyan}, {Teyssier}, \& {Gottloeber}}]{prunet08}
{Prunet} S., {Pichon} C., {Aubert} D., {Pogosyan} D., {Teyssier} R.,
  {Gottloeber} S., 2008, \apjs, 178, 179

\bibitem[{{Rasera} \& {Teyssier}(2006)}]{rasera06}
{Rasera} Y., {Teyssier} R., 2006, \aap, 445, 1

\bibitem[{{Rutkowski} {et~al}\mbox{.}(2012){Rutkowski}, {Cohen}, {Kaviraj},
  {O'Connell}, {Hathi}, {Windhorst}, {Ryan}, {Crockett}, {Yan}, {Kimble},
  {Silk}, {McCarthy}, {Koekemoer}, {Balick}, {Bond}, {Calzetti}, {Disney},
  {Dopita}, {Frogel}, {Hall}, {Holtzman}, {Paresce}, {Saha}, {Trauger},
  {Walker}, {Whitmore}, \& {Young}}]{rutkowski12}
{Rutkowski} M.~J. {et~al.}, 2012, \apjs, 199, 4

\bibitem[{{Sirko}(2005)}]{sirko05}
{Sirko} E., 2005, \apj, 634, 728

\bibitem[{{Stark} {et~al}\mbox{.}(2009){Stark}, {Ellis}, {Bunker}, {Bundy},
  {Targett}, {Benson}, \& {Lacy}}]{stark09}
{Stark} D.~P., {Ellis} R.~S., {Bunker} A., {Bundy} K., {Targett} T., {Benson}
  A., {Lacy} M., 2009, \apj, 697, 1493

\bibitem[{{Sutherland} \& {Dopita}(1993)}]{sutherland93}
{Sutherland} R.~S., {Dopita} M.~A., 1993, \apjs, 88, 253

\bibitem[{{Teyssier}(2002)}]{teyssier02}
{Teyssier} R., 2002, \aap, 385, 337

\bibitem[{{Teyssier} {et~al}\mbox{.}(2010){Teyssier}, {Chapon}, \&
  {Bournaud}}]{teyssier10}
{Teyssier} R., {Chapon} D., {Bournaud} F., 2010, \apjl, 720, L149

\bibitem[{{Thomas} {et~al}\mbox{.}(2005){Thomas}, {Maraston}, {Bender}, \&
  {Mendes de Oliveira}}]{thomas05}
{Thomas} D., {Maraston} C., {Bender} R., {Mendes de Oliveira} C., 2005, \apj,
  621, 673

\bibitem[{{Williams} {et~al}\mbox{.}(2009){Williams}, {Quadri}, {Franx}, {van
  Dokkum}, \& {Labb{\'e}}}]{williams09}
{Williams} R.~J., {Quadri} R.~F., {Franx} M., {van Dokkum} P., {Labb{\'e}} I.,
  2009, \apj, 691, 1879

\bibitem[{{Windhorst} {et~al}\mbox{.}(2011){Windhorst}, {Cohen}, {Hathi},
  {McCarthy}, {Ryan}, {Yan}, {Baldry}, {Driver}, {Frogel}, {Hill}, {Kelvin},
  {Koekemoer}, {Mechtley}, {O'Connell}, {Robotham}, {Rutkowski}, {Seibert},
  {Straughn}, {Tuffs}, {Balick}, {Bond}, {Bushouse}, {Calzetti}, {Crockett},
  {Disney}, {Dopita}, {Hall}, {Holtzman}, {Kaviraj}, {Kimble}, {MacKenty},
  {Mutchler}, {Paresce}, {Saha}, {Silk}, {Trauger}, {Walker}, {Whitmore}, \&
  {Young}}]{windhorst11}
{Windhorst} R.~A. {et~al.}, 2011, \apjs, 193, 27

\end{thebibliography}

\end{document}